\documentstyle[12pt,epsf]{article}
\newcommand {\bi} {\bibitem}
 \newcommand {\be} {\begin{equation}}
\newcommand {\bea} {\begin{eqnarray} \nonumber }
\newcommand {\ee} {\end{equation}}
\newcommand {\eea} {\end{eqnarray}}
 \newcommand {\eps} {\epsilon}
 \newcommand {\si} {\sigma}
\newcommand {\de} {\delta}
\newcommand {\De} {\Delta}
\newcommand {\ga} {\gamma}
\newcommand {\la} {\lambda}

 \newcommand {\al} {\alpha}
 
 \newcommand {\N} {{\cal N}}

\newcommand {\lan} {\langle}
\newcommand {\ran} {\rangle}

\newcommand {\cC}  {{\cal C}}

\newcommand {\cN}  {{\cal N}}

\newcommand {\cE}  {{\cal E}}
\newcommand {\cp} {\right)}
\newcommand {\ap} {\left(}

\newcommand {\for} {\ \ \ \mbox{for}\ \ }

\def \form#1 {eq. (\ref{#1}) }
\def \parziale#1#2  {{\partial {#1} \over \partial {#2}}}

\begin{document}
\title{Replica and Glasses}

\author{Giorgio Parisi\\
  {\small  Dipartimento di Fisica and INFN, Universit\`a di Roma}
   {\small {\em La Sapienza} }\\
{\small   P. A. Moro 2, 00185 Roma (Italy)}
{\small   \tt giorgio.parisi@roma1.infn.it }
}

\maketitle
\begin{abstract}
In these two lectures I review our theoretical understanding of spin glasses paying a particular 
attention to the basic physical ideas.  We introduce the replica method and we describe its 
probabilistic consequences (we stress the recently discovered importance of stochastic stability).  
We show that the replica method is not restricted to systems with quenched disorder.  We present the 
consequences on the dynamics of the system when it slows approaches equilibrium are presented: they 
are confirmed by large scale simulations, while we are still awaiting for a direct experimental 
verification.
\end{abstract}
\section{Introduction}

Many progresses have been done in the last years in the study of glass transitions using the replica 
approach.  The basic ideas are those of Kauzmann, Adams, Di Marzio and Gibbs, however the use of 
modern theoretical techniques gives a much better insight.

The basic hypothesis in this approach is that at low temperature the energy (or better the free 
energy) landscape contains an exponential number of minima.  The system may lives in one of this 
mimima, but the entropy of the system will get contribution also from the so called configurational 
entropy or complexity.

I will concentrate on three points in my lectures.
\begin{itemize}
\item
The study of soluble models in which the this scenario is exact.  The advantage of studying these 
model are the following:
\begin{itemize} 
\item You are able to obtain a precise formulation and define in a clear an precise manner the 
various concepts involved.
\item The resulting picture is much more complex that what you can be naively expect and a much
better inside is obtained,
\item You also obtain new qualitative predictions.
\item It is possible to positioning the mode coupling computations inside this scenario.  This is 
crucial in understand the utility and the limits of mode coupling approach.
\end{itemize}
\item
Model independent predictions for the generalization of the fluctuation dissipation theorem in 
off-equilibrium aging dynamics.  The fluctuation dissipation theorem does not hold when the system 
is reaching equilibrium and new relation are present, which are model independent: they have been 
tested in numerical simulation and we are eagerly waiting for an experimental test.
\item
The construction of first principle analytic computations which allows the computation of the 
property of glass forming quantity at all temperatures (i.e.  above and below the glass transition) 
by writing down and solving the appropriate integral equation for the correlation function.  

\end{itemize}
The tools used are mostly the replica theory and generalized two times mode coupling equations for 
the dynamics. 

In these lecture I will not present the detailed proof of the various result: I am 
trying to distill the conclusions of about one hundred papers.  I will mostly state the main 
results, in some cases sketch the proof and refer to the original literature for more details.
\section{The basic scenario}
In the nutshell many of the ideas I am going to present are not new: they are already in the 
original papers of Gibbs and Di Marzio.  However the comparison of glasses and generalized spin 
glasses, introduced in ref.  \cite {KWT} allow us to put these ideas in a  sharper form and to 
test them in numerical (and eventually real) experiments.  In this talk I will not discuss the 
theoretical basis under which this scenario has been derived (i.e.  the mathematical tool needed to 
derive the results stated for the generalized spin glasses) but I will concentrate the attention of 
the physical picture.

The basic ideas are quite simple \cite{CUKU,FRAPA,PARE,kurparvir,crisomtap,I,MONA}.  Let us consider 
a system of $N$ particles and let us assume that we can introduce a free energy functional $F[\rho]$ 
which depends on the density $\rho(x)$ and on the temperature.  We suppose that at sufficiently low 
temperature this functional has many minima (i.e.  the number of minima goes to infinity with the 
number ($N$) of particles).  Exactly at zero temperature these minima coincide with the mimima of 
the potential energy as function of the coordinates of the particles.  Let us label then by an index 
$\alpha$.  To each of them we can associate a free energy $F_\al$ and a free energy density $f_\al= 
F_\al/N$.

In this low temperature region we suppose that the total free energy of the system can be well 
approximated by the sum of the contributions to the free energy of each particular minimum:
\be
Z\equiv \exp(-\beta N f_{S}) =\sum_\al \exp(-\beta N f_\al).
\ee

When the number of minima is very high, it is convenient to introduce the function $\cN(f,T,N)$ 
which is the density of minima whose free energy is equal to $f$.  With this notation we can write 
the previous formula as
\be
Z= \int df \exp (-\beta N f) \cN(f,T,N).
\ee
In the region where $\cN$ is exponentially large we can write
\be
\cN(f,T,N) \approx \exp(N\Sigma(f,T)),\label{CON}
\ee
where the function $\Sigma$ is called the complexity or the configurational entropy (it is the 
contribution to the entropy coming from the existence of an exponentially large number of locally 
stable configurations).

The relation (\ref{CON}) is valid in the region $f_m(T)<f<f_M(T)$.  The minimum possible value of 
the 
free energy  is given by $f_m(T)$. Outside this region we have that $\cN(f,T)=0$.  It 
all cases known $\Sigma(f_m(T),T)=0$, and the function $\Sigma$ is continuous at $f_m$.

For large values of $N$ we can write
\be
\exp(-N \beta f_{S}) \approx \int_{f_m}^{f_M} df \exp (-N(\beta f- \Sigma(f,T)).\label{SUM}
\ee
We can thus use the saddle point method and  approximate the 
integral  with the integrand evaluated at its maximum.
We find that
\be
\beta f_{S}=\min_f\Phi(f) \equiv \beta f^* - \Sigma(f^*,T),
\ee
where
\be
\Phi(f)\equiv\beta f - \Sigma(f,T).
\ee
(This formula is quite similar to the well known homologous formula for the free energy ,i.e.  
$\beta 
f=\min_{E} (\beta E -S(E))$, where $S(E)$ is the entropy density as function of the energy density.)

If we call $f^*$ the 
value of $f$ which minimize $\Phi(f)$. we have two possibilities:
\begin{itemize}
\item
The minimum is inside the interval and it can be found as solution 
of the equation $\beta=\partial \Sigma/\partial f$.  In this case we have
\be
\beta \Phi=\beta f^* - \Sigma^*, \ \ \ \Sigma^*=\Sigma(f^*,T).
\ee
The system may stay in one of the many possible minima.  The number of minima at which 
is convenient for the system to stay is $\exp(N \Sigma ^*)$ .  The entropy of the system is thus 
the 
sum of the entropy of a typical minimum and of $\Sigma^*$, which is the contribution to the entropy 
coming from the exponential large number of microscopical configurations.

\item
The minimum is at the extreme value of  the range of variability of 
$f$.  We have that $f^*=f_m$ and $\Phi=f_m$.  In this case the contribution of the complexity to 
the 
free energy is zero.  The different states who contribute the the free energy have a difference in 
free energy density which is of order $N^{-1}$ (a difference in total free energy of order 1).
Sometimes we indicate the fact that the free energy is 
dominated by a few different minima by say the the replica symmetry is spontaneously broken 
\cite{mpv,parisibook2}.
\end{itemize}

Form this point of view the behaviour of the system will crucially depend on the free energy 
landscape \cite{ACP}, i.d.  the function $\Sigma(f,T)$, the distance among the minima, the height 
of 
the barriers among them...  Although our final task should be to compute the properties of the free 
energy landscape from the microscopic form of the Hamiltonian, we can tentatively assume that the 
landscape of fragile glasses is similar to that of some soluble long range models in presence of 
quenched disorder \cite {KWT,ACP}.

I cannot discuss here in details the rationale for this hypothesis; it is also clear that it cannot 
be exact and some differences should be present among the predictions of the mean field 
approximation and the real world.  Here I will present the scenario coming from mean field, 
stressing some of the predictions that should have a wider range of validity and comparing them 
with 
numerical simulations for fragile glasses.

 We can distinguish a few temperature regions.
\begin{itemize}
\item For $T>T_f$ the only minimum of the free energy functional is given by a constant density.
The system is obviously in the fluid phase.
\item For $T_f>T>T_D$ there is an exponentially large number of 
minima with a non-constant density $\rho(x)$ \cite{kurparvir,babumez}.  There are values of the 
free 
energy density such that the complexity $\Sigma$ is different from zero, however the contribution 
coming from these minima is higher that the one coming from the liquid solution with constant 
$\rho(x)$.
\item The most interesting situation happens in the region where $T_D>T>T_K$.  In this region the
free energy is still given the fluid solution with constant $\rho$ and at the same time the free 
energy is also given by the sum over the non trivial minima \cite{I,MONA}.

Although the thermodynamics is still given by the usual expressions of the liquid phase and final 
free energy is analytic at $T_D$, below this temperature the liquid phase correspond to a system 
which at each given moment may stay in one of the exponentially large number of minima.  It is 
extremely surprising that the free energy of the liquid can be written in this region as the sum of 
the contribution of the minima, according to formula (\ref{SUM}).  

The time to jump from one minimum to an other minimum is quite large: it is an activated process 
which is controlled by the height of the barriers which separate the different minima.  The 
correlation time will become very large below $T_D$ and for this region $T_D$ is called the 
dynamical transition point.  The correlation time (which should be proportional to the 
viscosity) should diverge at $T_{K}$.  The precise form of the this divergence is not well 
understood.
It is natural to suppose that we should get divergence of the form $\exp(A/(T-T_{K})^{\nu})$ for an 
appropriate value of $\nu$ \cite{VF}, whose reliable analytic computation is lacking \cite{KWT,PAK}.

The equilibrium complexity is different from zero (and it is a number of order 1) when the 
temperature is equal to $T_D$ and it decreases when the temperature decreases and it vanishes 
linearly at $T=T_K$.  At this temperature (the so called Kauzmann temperature) the entropy of a 
single minimum becomes to the total entropy and the contribution of the complexity to the total 
entropy vanishes.
\item
In the region where $T<T_K$ the free energy is dominated by the contribution of a few minima of the 
free energy having the lowest possible value.  Here the free energy is no more the analytic 
continuation of the free energy in the fluid phase.  A phase transition is present at $T_K$ and the 
specific heat is discontinuous here. 
\end{itemize}

\begin{figure}
\epsfxsize=250pt\epsffile{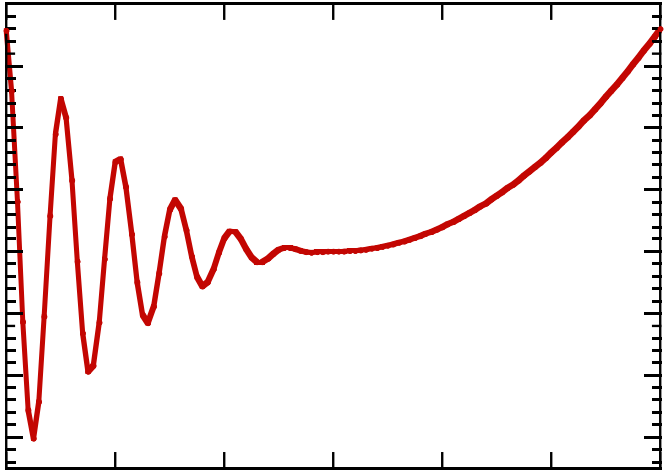}
\caption{The qualitative dependence of the free energy as function of the configuration space in 
the 
region relevant for the dynamical transition, i.e.  for $T<T_{D}$.  }
\end{figure}

The free energy landscape is quit usual; we can try do present the following pictorial 
interpretation, which is a rough simplification \cite{JL}.  At temperature higher than $T_{D}$ the 
system stays in a region of phase space which is quite flat and correspond of a minimum of the 
total 
free energy.  On the contrary below $T_{D}$ the phase space is similar to the one shown pictorial 
in 
fig.  1.  The region of maxima and minima is separated by the region without barriers by a large 
nearly flat region.  The minima in the region at the left are still present also when 
$T_{f}>T>T_{D}$, but they do not correspond to a global minimum.

At temperatures higher than $T_D$ the system at thermal equilibrium states in the right region.  
When the temperature reaches $T_D$ the system arrives in the flat region.  Here the potential is 
flat and this causes a more or less conventional Van Hove critical slowing down which is well 
described by the well known mode coupling theory \cite{vetro} (which is exact in the mean field 
approximation).  The mode coupling theory describes the critical slowing down which happens near 
$T_{D}$ \cite{BCKM}.

In the mean field approximation the height of the barriers separating the different minima is 
infinite and the temperature $T_D$ is sharply defined as the point where the correlation time 
diverge.  In the real world activated process (which are neglected in the mean field approximation 
and consequently in the mode coupling theory) have the effect of producing a finite (but large) 
correlation time also at $T_{D}$.  The precise meaning of the dynamical temperature beyond mean 
field approximation is discussed in details in \cite{FRAPA}

When the temperature is smaller that $T_D$ we must be more precise in describing the dynamics of 
the 
system.  Let us start from a very large system (of $N$ particles) at high temperature and let us 
gradually cool it. We find that it should go at equilibrium in the region with many minima.  
However 
coming from high free energy (from the right) it cannot enter in the region where are many maxima 
and minima if we wait a finite amount of time (the time to crosses the barriers diverges as $\exp 
(AN)$.  If we do not wait an exponentially large amount of time the system remains confined in the 
flat region.  In this case \cite{CUKU} the so called dynamical energy,
\be
E_D=\lim_{t \to \infty} \lim_{N \to \infty} E(t,N),
\ee
is higher that the equilibrium free energy.  The situation is described in fig.  2.

Below $T_D$ the system is trapped in metastable states when cooled.  The time needed to escape from 
these states diverges when $N$ goes to infinity in the mean field approach where activated 
processes 
are forbidden.  Of course the difference of the static and dynamic energy is an artifact of the 
mean 
field approximation if we take literarily the limit $t \to \infty$ in the previous equation because 
as matter of fact there are no metastable states with strictly infinite mean life.  However it 
correctly describe the situation on laboratory times, where metastable states are observed.

\begin{figure}
\epsfxsize=250pt\epsffile{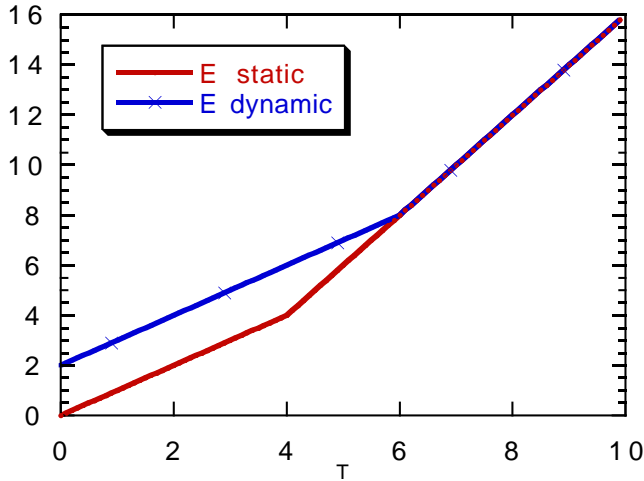}
\caption{The qualitative behaviour of the equilibrium and of the dynamical energy as function of 
the temperature.}
\end{figure}

In real systems, beyond the mean field approximation, the height of the barriers is finite also 
below $T_D$ and the mean life of the metastable states is finite, albeit very large.  Some 
mechanisms have been described who imply a divergence of the correlation time in real systems at 
the 
Kauzmann temperature \cite{KWT,FRAPA,PAK}.  The conventional glass temperature, i.e.  the 
temperature at which the microscopic correlation time becomes macroscopic (e.g.  of order of the 
minute) is between the two temperatures (i.e.  $T_D>T_G>T_K$).

It should be clear that in this framework the dynamical temperature $T_D$ is not so well defined 
and 
it correspond to a crossover region below it the dynamics is dominated by activated processes
\cite{FRAPA}.

In the mean field approximation there are very interesting phenomena that happen below $T_D$ when 
the system is cooled from the high temperature phase.  These phenomena are related to the fact that 
the system does not really go to an equilibrium configuration but wanders in the phase space never 
reaching equilibrium.  The phenomena are the following:
\begin{itemize}
\item
The energy approaches equilibrium slowly when the system is cooled from an high energy 
configuration.  In other words for large times we have
\be
E(t,T)=E_{D}(T)+B(T) t^{-\lambda(T)},
\ee
where the exponent $\la(T)$ does not vanish linearly at zero temperature as it should happens for 
an 
activated process.
\item
Aging is present, i.e.  the correlation functions and the response functions in the region of large 
time do depend on the story of the system \cite{B,POLI,FM}.
\item
In the region where aging is present the hypothesis at the basis of the fluctuation dissipation 
theorem are no more valid.  New generalized relations are satisfied \cite{FRARIE,MPRR}, which 
replace the fluctuation dissipation theorem.
\item
In the region where the diffusion constant is zero, a new phenomenon, logarithmic diffusion, is 
present.
\end{itemize}
All these phenomena are well known also in the case of spinodal decomposition but have a more 
general validity.
\section{The Random Energy Model}

\subsection{The definition of the model}

The Random Energy Model \cite{REM} is the simplest model for glassy systems.  It 
have various advantages: it is rather simple (its properties may be well 
understood with intuitive arguments, which may become fully rigorous) and 
display very interesting and new phenomena.

The Random Energy Model is defined as following.  There are $N$ Ising spins 
($\sigma_i$, $i=1,N$) which may take values $\pm1$; the total number of 
configurations is equal to $M\equiv 2^N$ and they can be identified by a label 
$k$ in the interval $1-M$.

Generally speaking the Hamiltonian of the system is given when we know the 
values of the energies $E_k$ for each of the $M$ configurations of the system.  
Usually one writes explicit expression for the energies as function of the 
configuration; on the contrary here we assume that the values of the $E_k$ are 
random, with a probability distribution $p(E)$ which is Gaussian:
\be
p(E) \propto \exp (-{E^2 \over 2 N}).\label{PRO}
\ee

The partition function is simply given by
\begin{eqnarray}
Z_\cE=\sum_{k=1,M} \exp (-\beta E_k)=\int \rho(E) \exp (-\beta E),\\
\rho(E)\equiv\sum_{k=1,M} \de(E-E_k).  \nonumber
\end{eqnarray}
   
The value of the partition function and of the free energy density 
$(f_\cE=-\ln(Z_\cE)/(N\beta )$) depends on $\cE$, i.e.  all the values of the 
energies $E_k$.  We would like to prove that when $N
\to \infty$ the dependance on $\cE$  of $f_\cE$ disappears with probability 
1.  If this happens, the most likely value of $f_\cE$ coincides with the average 
of $f_\cE$, where the average is done respect to all the possible values of the 
energy extracted with the probability distribution \form{PRO} .

The model is enough simple to be studied in great details; exact expressions can 
be derived also for finite $N$.  Here we sketch the results giving a 
plausibility argument.

\subsection{Equilibrium properties of the model}

The crucial observation is the following.  The probability of finding a 
configuration of energy $E$ is
\be
\N_0(E)\equiv 2^N \exp(-{E^2\over 2 N}) = \exp (N(\ln(2)- \frac12 e^2)),
\ee
where $e\equiv E/N$ is the energy density.  It is reasonable to assume (and it 
is confirmed by a detailed computation) that in the case of a generic system 
(with probability 1 when $N \to \infty$) no configurations are present in the 
region where $\N_0(E)<<1$, i.e.  for
\be e^2<e_c^2\equiv 2 \ln(2).  \ee 
We can thus 
write for the generic choice of the energies ($\cE$): 
\be \rho(E)\approx 
\N(E)\equiv N_0(E)\theta(E_c^2-E^2).  
\ee
 The partition function can be written 
as 
\be \int \N(E) \exp (-\beta E) .
\ee 
Evaluating the integral with the saddle 
point method one finds that in the high temperature region, i.e.  
\be 
\beta<\beta_c\equiv e_c^{-1}, 
\ee 
the internal energy density is simple given by $-\beta$.  This behaviour must 
end somewhere because we know that the energy is bounded form below also when 
$\beta \to \infty$.  Indeed in the low temperature region one finds that the 
integral is dominated by the boundary region $E\approx E_c$ and the energy 
density is exactly given by $-e_c$.

The entropy density is positive in the high temperature region, vanishes at 
$\beta_c$ and remains zero in the low temperature region.  It follows that in 
the high temperature region an exponentially large number of configurations 
contributes to the partition function, while in the low temperature region it is 
possible that the probability is concentrated on a finite number of 
configurations.

\subsection{Properties of the low temperature phase}

It is worthwhile to study in more details the structure of the configurations 
which mostly contribute to the partition function in the lower temperature 
phase.  At this end it is useful to sort the configurations with ascending 
energy.  We rename the configurations and we introduce new labels such that 
($E_k<E_i$ for $k<i$).

It is convenient to introduce the following quantity:
\be
w_k\equiv {\exp(-\beta E_k) \over Z}.
\ee
We have obviously that
\be
\sum_{k=1,2^N}w_k =1
\ee
A detailed computation shows \cite{REM,mpv,parisibook2} that in the low 
temperature region (i.e.  $\beta>\beta_c$) the previous sum is dominated by the 
first terms.  Indeed
\be
\sum_{k=1,L}  w_k =1-O(L^{-\la}), \ \ \ \la={1-m \over m},
\ee
where 
\be
m={T \over T_c}.
\ee
In the same region the sum in the following equation is convergent and its 
average value is given by
\be
\sum _{k=1,2^N}w_k^2=1-m.
\ee

Generally speaking one finds that one can introduce the quantities $F_k$ 
such that
\be
w_k \propto \exp(-\beta F_k)
\ee
Here the variables $F_k$ coincide with the total energy (not the energy 
density!) apart form an addictive constant. 
Their
probability distribution at the lower  end (which is the relevant region for 
thermodynamics in the low temperature region) can be approximated as
\be
P(F)\approx \exp (\beta m F).
\ee

In this model everything is clear: in the high temperature region the number of 
relevant configurations is infinite (as usual) and there is a transition to a 
low temperature region where only few configuration dominates. 

This phenomenon can be seen also in the following way.  We introduce a 
distance among two configurations $\al$ and $\ga$ as
\be
d^2(\al,\ga) \equiv {\sum_{i=1,N} (\si^\al_i -\si^\ga_i)^2 \over 2 
n}.\label{DISTANZA}
\ee
Sometimes it is convenient to introduce also the overlap $q$defined as
\be
q(\al,\ga)\equiv {\sum_{i=1,N} \si^\al_i \si^\ga_i \over 2 N}=1-d^2(\al,\ga).
\ee
The distance squared is normalized in such a way that it spans the interval 
$0-2$. It is equal to
\begin{itemize}
\item
 0, if the two configuration are equal ($q=1$).
 \item 
 1, if the configuration are orthogonal ($q=0$).
 \item
 2, if $\si^\al_i=-\si^\ga_i$ ($q=-1$).
 \end{itemize}
 
It is convenient to introduce the function $Q(d)$ and $P(q)$, i.e.  the 
probability that two equilibrium configurations are at distance $d$ or 
overlap $q$ respectively.  We find 
\begin{itemize} 
\item For $T>T_c$ 
\be 
Q(d)=\delta(d-1), \ \ \ P(d)=\de(q).  
\ee 
\item For $T<T_c$ 
\be Q(d)=(1-A) 
\delta(d-1)+A\delta(d), \ \ \ P(d)=(1-A) \delta(d)+A\delta(q-1).  
\ee
where $A$ is equal to $\sum _{k=1,2^N}w_k^2$.  The average of $A$ over the 
different realizations of system is equal to $1-m$.
\end{itemize}

 As soon as we enter in the low temperature region, the 
probability of finding two equal configurations is not zero.  The 
transition is quite strange from the thermodynamic point of view.
\begin{itemize}
\item
It looks like a {\sl second} order transition because there is no latent heat.
It is characterized by a jump in the specific heat (which decreases going toward 
low temperatures).
\item
It looks like a {\sl first} order transition.  There are no divergent 
susceptibilities coming form above of below (which beyond mean field theory 
should imply no divergent correlation length).  Moreover the minimum value 
of $d$ jumps discontinuously (from 1 to 0).

\item If we consider a system composed by two replicas ($\si^1$ and $\si^2$)  
\cite{kurparvir} and
we write the Hamiltonian
\be
H(\si^1,\si^2)=H(\si^1)+H(\si^2)+N \eps d^2(\si_1,\si^2)
\ee
the thermodynamics is equal to that of the previous model (apart a factor 
2) for $\eps=0$, but we find a real first order thermodynamic transition, 
with a discontinuity in the internal energy, as soon as $\eps>0$.  The case 
$\eps=0$ is thus the limiting case of real first order transitions.
\end{itemize}
 
These strange characteristics can be summarized by saying that the 
transition is of order one and half, because it share some characteristics 
with both the first order and the second order transitions.
 
It impressive to note that the thermodynamic behaviour of real glasses near 
$T_c$ is very similar to the order one and half transition of REM. We will sea 
later that this behaviour is typical of the mean field approximation to glassy 
systems.

\subsection{Dynamical properties of the model}
The dynamical properties of the model can be easily investigated in a 
qualitative way. Interesting 
behavior is present in the region where the value of $N$ is large with respect 
to the time. Different results will be obtained for different definition of the 
dynamics if we consider some rather artificial form of the dynamics.

Let us first consider a single spin flip dynamics.  In other words we assume 
that in a microscopic time scale scale, which for simplicity we consider of 
order unit, the system explore all the configurations which differ from the 
original one by a single spin flip and goes to one of them (or remain in the 
original one) with probability which is proportional to $\exp (-\beta H)$.  This 
behaviour is typical of many dynamical process, like Glauber dynamics, monte 
Carlo, heath bath.

In this dynamical process each configuration $C$ has $N$ nearby configurations 
$C'$ to explore.  The energies of the configurations $C'$ are uncorrelated to 
the energy of $C$, so that they would be of order one in most of the case.  The 
lowest energy of the configurations $C'$ would be of order $-(N\ln(N))^{1/2}$, 
the corresponding energy density ($-(\ln(N)/N)^{1/2}$) vanishes in the large $N$ 
limit.

If the configuration $C$ has an energy density $e$ less that zero, but 
greater than the equilibrium energy, the time needed to do a transition to 
a configuration of lower energy will be, with probability one,
exponentially large.

 For 
short times a configuration of energy $e$ will be completely frozen.  Only 
at  larger times it may jump to a 
typical configuration of energy zero.  At later times different scenarios 
are possible: the configuration comes back to the original configuration of 
of energy $e$ or, after some wandering int the region of configurations of 
energy density $\approx 0$, it fells in an other deep configuration of 
energy $e'$.  A computation of the probabilities for these different 
possibilities has not yet been done, although it should not too difficult.

The conclusions of this analysis are quite simple. 
\begin{itemize}
\item
If we start from a random configuration, after a time which is finite when 
$N\to \infty$, the system goes to a configuration whose energy is of order 
$-\ln(N)^{1/2}$ and stops there.
\item If we start from a random configuration and we study the system at
exponentially large times the system will reach an energy density which may 
be different from the original one.
\end{itemize}

\section{Models with partially correlated energy}
\subsection{The definition of the models}
The random energy model (REM) is rather unrealistic in that it predicts that the 
energy is completely upset by a single spin flip.  This feature can be 
eliminated by considering a more refined model, the so called $p$-spins models 
\cite{GROMEZ,GARDNER} , in which the energies of nearby configurations are also 
nearby.  We could say that energy density (as function of the configurations) 
is not a continuous function in the REM, while it is continuous in the $p$-spins 
models, in the topology induced by the distance (\form{DISTANZA} ).  In this new 
case some of the essential properties of the REM are valid, but new features are 
present.

The Hamiltonian we consider depends on some control variables $J$, which have a 
Gaussian distribution and play the same role of the random energies of the REM 
and by the spin variable $\si$. For $p=1,2,3$ the Hamiltonian is respectively
\begin{eqnarray}
H^1_J(\si)= \sum_{i=1,N} J_i \si_i\\
H^2_J(\si)= \sum_{i,k=1,N}' J_{i,k} \si_i \si_k\\
H^3_J(\si)= \sum_{i,k,l=1,N}' J_{i,k,l} \si_i \si_k \si_l \nonumber
\end{eqnarray}
where the primed sum indicates that all the indices are different. The variables
$J$ must have a variance of $O(N^{(1-p)/2})$ in order to have a non trivial 
thermodynamical limit.

It is possible to prove by am explicit computation that if we send first $N \to 
\infty$ and later $p \to \infty$, one recover the REM.  Indeed the energy 
differences corresponding to one spin flip are of order $p$ for large $p$ 
(they ar order $N$ in the $REM$, so that in the limit $p \to \infty$ the 
energies in nearby configurations become uncorrelated and the REM is 
recovered.

\subsection{Equilibrium properties of the models}

The main new property of the model is the correlation of energies.  This fact 
implies that if $C$ is a typical equilibrium configuration, all the 
configurations which differ from it by a finite number of spin flips will also 
have a finite energy.  The equilibrium configurations are no more isolated (as 
in REM), but they belongs to valleys, such that the entropy restricted to a 
single valley is proportional to $N$ and it is and extensive quantity.

The thermodynamical properties at equilibrium can be computed using the re\-plica me\-thod 
\cite{mpv,parisibook2} .  Let us discuss firstly what happen for $p>2$ In the simplest version of 
this method
\cite{GROMEZ,GARDNER} one introduces the typical overlap of two configurations 
inside the same valley (sometimes denoted by $q_{EA}$).  Something must be said 
about the distribution of the valleys.  Only those which have minimum free 
energy are relevant for the thermodynamics.  One finds that these valleys have 
zero overlap and have the same distribution of free energy as in the REM
\be
P(F)\propto \exp(\beta m (F-F_0)).
\ee
Indeed the average value of the free energy can be written in a self consistent 
way as function of $m$ and $q$ ($f(q,m)$) and the value of these two parameters 
can be found as the solution of the stationarity equations:
\be
\parziale{f}{m} =\parziale{f}{q} =0.
\ee

The quantity $q$ (which would be 1 in the REM) is here of order $1-\exp(-A\beta 
p)$ for large $p$, while the parameter $m$ has the same dependence of the 
temperature as in the REM, i.e.  1 at the critical temperature, and a linear 
behaviour al low temperature.  The only difference is that $m$ is no more 
strictly linear as function of the temperature.

The thermodynamical properties of the model are the same as is the REM: a 
discontinuity in the specific heat, with no divergent susceptibilities.
Let us first recall the description of these glassy systems at equilibrium according to the 
predictions of the replica theory.  We consider a systems and we denote by $\cal C$ a generic 
configuration of the system.  For simplicity we will assume that there are no symmetry in the 
Hamiltonian, in presence of symmetries the arguments must be slightly modified.  It is useful to 
introduce an overlap $q(\cC,\cC')$.  There are many ways in which an overlap can be defined; for 
example in spin system we could define
\be
q={\sum_{i=1,N}\si_{i}\tau_{i}\over N},
\ee
$N$ being the total number of spins or particles and $\si$ and $\tau$ are the two spin 
configurations.  In a liquid a possibility is given by
\be
 q={\sum_{i=1,N}\sum_{k=1,N}f(x(i)-y(k))\over N},
\ee
where $f$ is a function which decays in a fast way at large distances and is substantially different 
from zero only at distances smaller that the interatomic distance ($x$ and $y$ are the two 
configurations of the system).

In the high temperature phase for very large values of $N$  the probability distribution 
of the overlap ($P_{N}(q)$) is given by
\be
P_{N}(q)\approx \delta (q-q^{*}).
\ee
In the low temperature phase $P_{N}(q)$ depends on $N$ (and on the quenched disorder, if it is 
present). When we average over $N$ we find a function $P(q)$ which is not a simple delta 
function.  In all known case one finds that
\be
P(q)=a_{m}\delta(q-q_{m})+a_{M}\delta(q-q_{M})+p(q),
\ee
where the function $p(q)$ does not contains delta 
function and its support is in the interval $[q_{m},q_{M}]$.

The non triviality of the function $P(q)$ (i.e.  the fact that $P(q)$ is not a single delta 
function and consequently $q$ is an intensive fluctuating quantity) is related to the existence of 
many  different equilibrium states.  Moreover the function $P_{N}(q)$ changes with $N$ and its 
statistical properties (i.e the probability of getting a given function $P_{N}(q)$) can be 
analytically computed \cite{mpv,BAPA}.

In this equilibrium description a crucial role is given by the function $x(q)$ defined
as
\be
x(q)=\int_{q_{m}}^{q}P(q')dq'\ .
\ee

In the simplest case the function $p(q)$ is equal to zero. i.e the function $P(q)$ has only two delta 
functions without the smooth part.  In this case, which correspond to one step replica symmetry 
breaking, there are many equilibrium states, labeled by $\al$, and the overlaps among two generic 
configurations of the same state and of two different states are respectively $q_{M}$ and $q_{m}$.  
The probability of finding a state with total free energy $f$ is proportional to
\be
\exp \left( m\beta(f- f_{R})\right), \label{REM}
\ee
where $f_{R}$ is a reference free energy and $m$ is the value of $x(q)$ in the interval 
$[q_{m},q_{M}]$.

In the more complicated situation where the function $p(q)$ is non zero, couples of different states 
may have different values of the overlaps.  The joint probability distribution of the states and of 
the overlaps can be described by formulae similar to eq.  (\ref{REM}), but more complex \cite{mpv}.

Although these predictions are quite clear, it is not so simple to test them for many reasons:
\begin{itemize}
\item They are valid  at thermal equilibrium,  a condition that is very difficult
to reach for this kind of systems.
\item Experimentally is extremely difficult to measure the values of the microscopic variables, i.e 
all the spins of the system at a given moment.  These measurements can be done only in numerical 
simulations, where the observation time cannot be very large.
\end{itemize}

A very important progress has been done when it was discovered \cite{CUKU} that the function $X$, 
which describes the violations of the fluctuation dissipation theorem, is equal to the function $x$ 
which is relevant for the statics.  This equality is very interesting because function $X(C)$ can be 
measured relatively easily in off-equilibrium simulations
\cite{FRARIE}.

The temperature dependence of the function $X(C)$ (or equivalently $x(q)$) is interesting 
also because rather different systems can be classified in the same universality class according to 
the behaviour of this function.  It has been conjectured long time ago that the equilibrium 
properties of glasses are in the same universality class of some simple generalized spin glass 
models
\cite{KWT,PARI}.

\subsection{The Free energy landscape}
It would be interesting to characterize better  the free energy landscape of 
the model, especially in order to understand the dynamics. Indeed we have 
already seen that in the REM the system could be trapped in metastable 
configurations. Here the situation is more complicated. Although the word {\sl 
valley} has a strong intuitive appeal, we must first define 
what a valley is in a more precise way.

There are two different (hopefully equivalent) definitions of a valley:
\begin{itemize}
\item
A valley is a region of configuration space separated by the rest of the 
configuration space by free energy barriers which diverge when $N\to\infty$.  
More precisely the system, in order to go outside a valley by moving one spin at 
once, must cross a region where the free energy is higher that of the valley by 
a factor which goes to infinity with $N$.
\item
A valley is a region of configuration space in which the system remains for time 
which goes to infinity with $N$.
\end{itemize}
The rationale for assuming that the two definitions are equivalent is the 
following.  We expect that for any reasonable dynamics in which the systems 
evolves in a continuous way (i.e.  one spin flip at time), when it goes from a 
valley to an other valley, the system must cross a configuration of higher free energy 
and therefore the time for escape from a valley is given by
\be
\tau \simeq \tau_0 \exp (\beta \Delta F)
\ee
where $\Delta F$ is the free energy barrier.

It is crucial to realize that in  infinite range models there can be valley which 
have an energy density higher that that of equilibrium states. This phenomenon 
is definitely not present in short range models. No metastable states with 
infinite mean life do exist in nature. 

Indeed let us suppose that the system may stay in phase (or valleys) which we 
denote as $A$ and $B$.  If the free energy density of $B$ is higher than that of 
$A$, the system can go from $B$ to $A$ in a progressive way, by forming a bubble 
of radius $R$ of phase $A$ inside phase $B$.  If the surface tension among phase 
$A$ and $B$ is finite, has happens in any short range model, for large $R$ the 
volume term will dominate the free energy difference among the pure phase $B$ 
and phase $B$ with a bubble of $A$ of radius $R$.  This difference is thus 
negative at large $R$, it maximum will thus be finite.  In the nutshell a finite 
amount of free energy in needed in order to form a  seed of phase $A$ starting 
from which the spontaneous formation of phase $A$ will start. For example, if 
we take a mixture of $H_2$ and $O_2$ at room temperature, the probability of a 
spontaneous temperature fluctuation  in a small region of the sample, 
which lead to later ignition and eventually to the explosion of the whole 
sample, is greater than zero (albeit quite a small number), and obviously it does not go 
to zero when the volume goes to infinity.

We have two possibilities open in positioning this mean field theory prediction 
of existence of real metastable states:
\begin{itemize}
	\item  We consider the presence of these metastable state with {\sl infinite} 
	mean life an artefact of the mean field approximation and we do not pay 
	attention to them.

\item We notice that in the real systems there are metastable states with very 
large (e.g.  much greater than one year) mean life.  We consider the {\sl 
infinite} time metastable states of the mean field approximation as precursors 
of these {\sl finite} states.  We hope (with reasons) that the corrections to 
the mean field approximation will give a finite (but large) mean life to 
these states (how this can happen will be discussed in the next section).
\end{itemize}

Here we suppose that the second possibility is the most interesting and we 
proceed with the study of the system in the mean field approximation.  The 
strategy for investigate the properties of these metastable states consists in 
considering systems with $R$ replicas (two or more) of the same system with 
Hamiltonian given by
\be
\beta H = \sum_{r=1,R}\beta_r H(\si^r) + \sum_{r,s=1,R}\eps_{r,s} q_{r,s}.
\ee
Different replicas may stay at different temperature. The quantities $\eps$ are 
just Legendre multipliers needed to enforce  specific value of the the overlaps 
$q$. In this way (let us consider for simplicity the case where all temperature 
are zero) we find (after a Legendre transform) a free energy density as function 
of the $q$. 

Let us consider for simplicity the case where we set
\be
q_{1,r}= q \for r=2, R
\ee
($q_{r,r}$ is identical equal to 1) and the others $q$ are left free \cite 
{kurparvir} .  A simple computation show the final free energy density that we obtain (let us call 
$f_R(q)$ is given by
\begin{eqnarray}
f_R(q) - R f= - \lim_{N \to \infty} {\ln
 \ap \sum_\si \ap \sum_\tau \delta(q(\si,\tau)-q) \cp ^{(R-1)} \cp \over \beta N}\equiv \nonumber\\
- \lim_{N \to \infty} {\ln <P_\si(q)^{(R-1)}> \over \beta N},
\end{eqnarray}
where $f$ is the unconstrained free energy.

A particular case, which is very interesting, is given by the limit $R \to 1$ 
\cite{ PAK,FP} :
\be
W(q)\equiv \lim_{N \to \infty} {\ln (P_\si(q))\over \beta N}=
\parziale{(f_R(q) - R f)}{R} |_{R=1}
 \ee

The potential $W(q)$ has usually a minimum at $q=0$, where $W(0)=0$. It may have a secondary 
minimum at $q=q_D$. We can have three quite different situations
\begin{itemize}
\item $W(q_D)=0$.  This happens in the low temperature region, below $T_c$, 
where we can put two replicas both at overlap $0$ and at overlap $q_{EA}$ 
without paying any prize in free energy.  In this case $q_D=q_{EA}$.

\item $W(q_D)>0$.  This happens in an intermediate temperature region, above 
$T_c$, but below $T_D$, where we can put one replica $\si$ at equilibrium and 
have the second replica $\tau$ in a valley near it. It happens that the 
internal energy of both the $\si$ configuration (by construction) and of the 
$\tau$ configuration are equal to the equilibrium one. However the number of 
valley is exponentially large so that the free energy a single  valley will be 
smaller. One finds in this way that $W(q_D)>0$ is given by
\be
W(q_D)= {\ln \N _e \over N}
\ee
where $ \N _e $ is the average number of the 
valleys having the equilibrium energy \cite{MONA,PP} .

	\item  At $T>T_D$ the potential $W(q)$ has only the minimum at $q=0$. The 
	quantity $q_D$ cannot be defined and no valley 
	with the equilibrium energy are present. This is more or less the definition 
	of the dynamical transition temperature $T_D$. A more careful analysis shows that 
	for $T_D<T<T_V$ there are still valleys with energy {\sl less} than the equilibrium 
	one, but these valleys cover a so small region of phase space that they are not 
	relevant for equilibrium physics.
\end{itemize}

It is also possible to study the properties of the free energy for $R\ne 1$ we 
can force the $\si$ configuration to be not an equilibrium one and in this way 
we control the properties of  the valleys having an energy different than the 
equilibrium one.
This method can give rather detailed information on the free energy landscape, 
which I do not have time to discuss in details and which have not yet yet fully 
studied.

The most interesting result is that for $T<T_D$ the entropy of the system can be 
written as
\be
S= S_V + W
\ee
where $S_V$ is the entropy inside a valley and $W$ is the configurational 
entropy, or complexity, i.e. the term due to the existence of an exponentially 
large number of states. The $W$ contribution vanishes at $T_c$ and becomes 
exactly equal to zero for $T<T_c$ \cite{KWT} .

In the REM limit ($p\to \infty$) the temperature $T_D$ goes to infinity.  
In this limit the third region does not exist.  Therefore the dynamical 
transition is a new feature which is not present in the REM.

\section{Static and dynamic properties}The simplest way to study the dynamics of the problem is to 
consider a
system which evolves according to Langevin equation of to some sort of 
Glauber dynamics.  For example we can suppose that:
\be
{d \si_i \over dt} = -{\de H \over \si_i} +\eta_i(t),
\ee
where $\eta$ is an appropriate white noise.

In this model is convenient to introduce the
single site
correlation function ($C$) and the response ($G$) function of the times.
One finds that they can be defined as 
\begin{eqnarray}
C(t_1,t_2)=<\si_i(t_1) \si_i(t_2)>\\
G(t_1,t_2)=<{\de \si_i(t_1) \over h_i(t_2)}>
\end{eqnarray}
where $h(t)$ is an external magnetic field. 

If the systems is at equilibrium (or in a metastable state), the 
correlation  and the response functions will depend only on the time 
difference. If the system is out of equilibrium these functions will 
depend in a non trivial way from both the arguments.

In both cases one can write down closed equations for the correlation 
functions in the case where the size $N$ goes to infinity at fixed times 
\cite{CUKU} .  These 
equations have a rather complex structure.  We could discuss two different 
regimes:
\begin{enumerate}
	\item  We start at time zero from an equilibrium configuration.

	\item  We  start at time zero from a random configuration.

\end{enumerate}
In the first case we find that, if we approach the dynamical temperature 
from above, the correlation time diverges a a power of $T-T_D$, and the 
usual analysis of the mode coupling theory can be done in this region. Mode 
coupling theory is essentially correct in this region.

In the second case one finds that below the dynamical transition, the energy 
does not go anymore to the equilibrium value, but it goes to an higher value, 
which in some cases can be computed analytically.

The phenomelogy is rather complex, the aging properties of the systems are 
particularly interesting, but we cannot discuss them for lack of space. It is 
interesting to note that the mode coupling theory become exact in the mean 
field theory and describes what happens nearby the dynamical phase transition 
at $T_D$, which is a temperature  higher that the equilibrium transition 
temperature at $T_c$.

Here we would like to understand the dynamical properties of the model starting 
from results which we have already derived on the free energy landscape without 
having to compute explicitly the solution of the equation of motion for the 
correlations $C$ and $G$. This can be done by assuming the this very slow 
dynamics is an activated process dominated by barrier crossing and that the 
height of the barriers can be computed using the method of the previous 
sections. 

Some of the question we ask are the following:
\begin{itemize}
\item How long does a system remains  in the same 
valley?
\item If the system is out equilibrium at time 0, which is the asymptotic value of 
the energy at large times?
\end{itemize}
 The results are 
similar to those obtained by the explicit dynamical analysis of the previous 
section.  We obtain also more detailed information on the region where the time 
scale is exponentially large.  This region can be studied only with very great 
difficulties using the dynamical equations.

In the first case (i.e.the system start from a thermalized configuration)
below $T_D$, the system is confined to a valley.
A first estimate of the time needed to escape from a valley can be obtained as 
follows. We introduce the parameter 
\be
q(t)\equiv {\sum_{i=1,N} \si_i(0) \si_i(t) \over N}.
\ee

We assume that on the large time scale the evolution of $q$ is more or less the 
same of a system with only one degree of freedom with potential $W(q)$. This 
assumption is not completely correct, but it is likely to be enough to give a 
first estimate (to be refined later) of the escape time.
Therefore it is reasonable to assume that in a first approximation
the the time needed to escape from the valley is given by 
\be
\tau=\tau_0 \exp(N \beta \De W),
\ee
where $\De W$ is the difference in free energy among the minimum at $q_D$ 
and the maximum al lower values of $q$.

For large time (but not exponentially divergent with $N$) we have that
\be
\lim_{t \to \infty}q(t) = q_D.
\ee
Eventually the system will escape from a valley for times greater that 
$\tau$.

In the other situation (the system is originally out equilibrium) things are 
more complicated.  Metastable valleys of high energies do exist, however it is 
not clear that a system cooled from an high temperature region must be trapped 
in the valley with high energy.  Indeed it will be trapped in the valley with 
largest attraction domain, but the properties of the valley cannot be found if 
we do not use the dynamics in an explicit way.  Some educated conjectures can be 
done, but the question has never been investigated in details.  The problem is 
not easy, because the system is strongly out of equilibrium.

A better understanding comes if we consider the case in which the system is 
slowly cooled, from high to low temperature. We consider here that case of 
slow, not ultra slow, cooling, i.e. the time scale if fixed when $N$ goes 
to infinity. In this case it is reasonable to assume that the system frozen 
in a valley at the dynamical temperature and we have to follow the energy 
of that valley when we cool. That can be done by using the formulation with 
different replicas at different temperatures. In this way one finds that 
the system is frozen in a configuration which is near to the configuration 
at the dynamical temperature.

A detailed comparison of the results for the energy of the metastable states 
obtained by doing different assumption is 
still lacking, but is should be not too difficult.

\section{Systems without quenched disorder}
Apparently the previous discussions are restricted to systems where  quenched disorder is present.  
The requirement of quenched disorder would limit ourselves in the applications of the replica method 
and one would cut all those systems, like glasses, which have a translational invariant Hamiltonian 
and where no quenched disorder is present.

This prejudice (on the need of quenched disorder( was so strong that it took a few years to realize 
that the replica method can also applied to system without quenched disorder.

There are many facts that clearly indicate the possibility of applying the replica methods to 
non-random systems.
\begin{itemize}
\item In the infinite range case there are pairs of systems with Hamiltonian respectively $H_{Q}$ and
$H$, where $H_{Q}$ contains quenched disorder and no disorder is present in $H$, such the high 
temperature expansion for the two systems coincide \cite{MPR,FH}.  It is natural to suppose that the 
free energies of the two system are identical at all temperatures, so that replica symmetry breaking 
can be applied to both.

\item It is possible to look for replica symmetry breaking in the expression for the free 
energy systems without disorder (e.g.  soft spheres) inside a given approximation (e.g.  
hypernetted chain) and find out that replica symmetry breaks at low temperature \cite{MP}.

\item In the replica method we can introduce coupled replica potentials \cite{PV} in order to
characterize the phase space of the system and these potentials can also be computed for non-random 
systems, obtaining the same results as for random systems \cite{FP}.  This may be done analytically 
for soft spheres using the same approximation as before \cite{CFP}.

\item It is now clear that the replica method may be applied any stochastic stable system.  Indeed
stochastic stable systems are the limit of disorder systems where the replica method can be 
applied without problems.  Systems without quenched disorder may be stochastically stable if the 
free energy is computed using the Cesareo limit (i.e.  averaging over $N$).
\end{itemize}

This new perspective allows us to use the replica method in systems quite different from the usual 
one, e.g.  structural glasses, where no quenched disorder is present.
\section{Fluctuation and dissipation relations in aging dynamics}
\subsection{Theoretical considerations}

When we suddenly decrease the temperature in an Hamiltonian system, many interesting phenomena 
happen if the initial and the final temperatures correspond to different phases.  When the low 
temperature phase can be characterized by a simple order parameter (e.g.  the magnetization for 
ferromagnets) we find the familiar phenomenon of spinodal decomposition characterized by growing 
clusters of different phases.  There is a dynamical correlation length (i.e.  the size of the 
clusters), which increases as a power of the time $t$ after the quench, and the energy approaches 
equilibrium with power like corrections (e.g.  $E(t)\approx E_{\infty}+At^{-1/2}$).  In this region 
aging phenomena are also present \cite{B}.

The situation is more intriguing in the case of structural glasses and spin glasses where the low 
temperature phase cannot be characterized in term of a simple order parameter.  Remarkable 
progresses in understanding the off-equilibrium dynamics and its relations to the equilibrium 
properties has been done by noticing that a crucial off-equilibrium feature is the presence of 
deviations from the well known {\sl equilibrium} fluctuation-dissipation relations.  On the basis of 
analytic results for soluble models it has been conjectured that we can define a function $X(C)$, 
$C$ being an autocorrelation function at different times \cite{CUKU,FM,BCKM}.  This function 
characterizes the violations of the fluctuation-dissipation theorem (which is correct only at 
equilibrium).  It is remarkable that (at least in the case of spin glasses) the function $X(C)$ is 
equal to the function $x(q)$ ($q$ being the overlap of two spin configurations) which plays a 
central role in the equilibrium computation of the free energy \cite{mpv}.

Let us be more precise.  We concentrate our attention on a quantity $A(t)$.  We suppose that the 
system starts at time $t=0$ from an initial condition and subsequently it remains at a fixed 
temperature $T$.  If the initial configuration is at equilibrium at a temperature $T'>T$, we observe 
an off-equilibrium behaviour.  We can define a correlation function
\be
C(t,t_{w}) \equiv \lan A(t_{w}) A(t+t_{w})\ran
\ee
and the response function
\be
G(t,t_{w}) \equiv \frac{ \de \lan A(t+t_{w})\ran}{\de \eps(t_{w})}{\Biggr |}_{\eps=0},
\ee
where we are considering the evolution in presence of a time dependent Hamiltonian in which we have
added the term
$ \int dt \eps(t) A(t) $.
 
The usual equilibrium fluctuation-dissipation theorem (FDT) tells us that
\be G^{eq}(t)= - \beta \frac{d C^{eq}(t)}{ dt}, \ee
where
\be
G^{eq}(t)=\lim_{t_w \to \infty} G(t,t_w), \ \ C^{eq}(t)=\lim_{t_w \to \infty} C(t,t_w).
\ee

It is convenient to define the integrated response:
\be
R(t,t_{w})=\int_{0}^{t} d\tau G(t-\tau,t_{w}+\tau),\ \ R^{eq}(t)=\lim_{t_w \to \infty} R(t,t_w),
\ee
$R(t,t_{w})$ is the response of the system at time $t+t_{w}$ to a field acting for a time $t$ 
starting at $t_{w}$.  The usual FDT relation becomes
\be
R^{eq}(t)= \beta (C^{eq}(t)-C^{eq}(0)).
\ee

\begin{figure}
\epsfysize=250pt
\epsffile{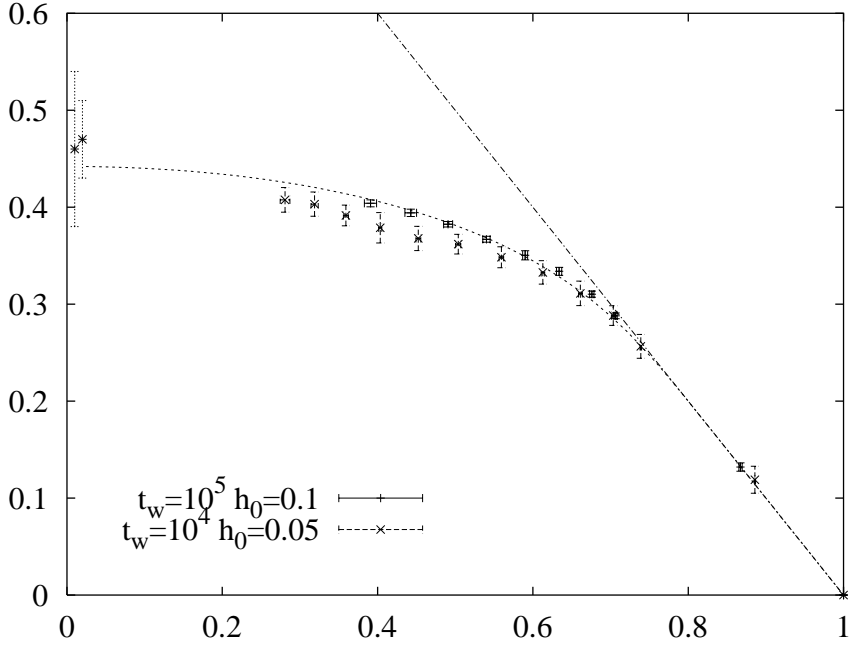}
\caption{The response $ R$ times $T$ versus $ C$ at $T=0.7$ for the three
dimensional Ising spin glass [10].  The curve is the prediction for function $R(C)$ obtained from the 
equilibrium data.  The straight line is the FDT prediction.  We have plotted the data of the two 
runs: $t_w=10^5$, and $t_w=10^4$.}
\protect\label{fig:gfdt}
\end{figure}

The off-equilibrium fluctuation-dissipation relation \cite{CUKU} states that the 
response function and the correlation function satisfy the following relation for large $t_w$:
\be
R(t,t_w)\approx \beta \int_{C(t,t_w)}^{C(0,t_{w})}X(C) dC.  \label{OFDR}
\ee
If we plot $R$ versus $\beta C$ for large $t_{w}$ the data collapse on the same 
universal curve and the slope of that curve is $-X(C)$.  The function $X(C)$ is system dependent and 
its form tells us  interesting information. in the case of three dimensional spin glasses.

We must distinguish two regions:
\begin{itemize}
\item A short time region where $X(C)=1$ (the so called FDT region) and $C$ belongs to the interval
$I$ 
(i.e. $C_1<C<C_2$.).

\item  A large time region (usually $t=O(t_w)$) where 
$C\notin I$ and $X(C)<1$.  In the same region the correlation function often satisfies an aging 
relation i.e $C(t,t_w)$ depends only on the ration $s \equiv t/t_{w}$ in the region where both $t$ 
and $t_{w}$ are large: $C(t,t_w)\approx C^{a}(t/t_{w})$.
\end{itemize}

In the simplest non trivial case, i.e.  one step replica symmetry breaking \cite{mpv,PARI} , the 
function $X(C)$ is piecewise constant, i.e.
\be
X(C)= m \for C \in I,\ \ X(C)= 1 \for C \notin I \label{ONESTEP}.
\ee
One step replica symmetry breaking for glasses has been conjectured in ref.  \cite{KWT}.

In all known cases in which one step replica symmetry holds, the quantity $m$ vanishes linearly with 
the temperature at small temperatures.  It often happens that $m=1$ at $T=T_{c}$ and $m(T)$ is 
roughly linear in the whole temperature range.  

\subsection{Stochastic stability}

Stochastic stability is a property which is valid in the mean field approximation; it is however 
possible to conjecture that is valid in general also for short range models.  It has been 
introduced quite recently \cite{GUERRA,AI,SOL,FMPP} and strong progresses have been done on the study 
of its consequences.

In order to decide if a system with Hamiltonian $H$ is stochastically stable, we have to consider 
the free energy of an auxiliary system having the following Hamiltonian:
\be
H+\eps^{1/2}H_{R}.
\ee
If the average (with respect to $H_{R}$) free energy is a differentiable function of $\eps$ (and the 
limit volume going to infinity commutes with the derivative with respect to $\eps$), for a generic 
choice of the random perturbation $H_{R}$ inside a given class and $\eps$ near to zero, the system is 
stochastically stable.
 
The definition of stochastic stability may depend on the class of random perturbations we 
consider.  Quite often it is convenient to chose as a random perturbation an infinite range 
Hamiltonian, e.g.
\be
H_{R}=\sum_{i,k,l}J_{i,k,l} \si_{i} \si_{k} \si_{l} \label{SSS}
\ee
where sum runs over all the $N$ points of the system and the $J$'s are random variables with 
variance $1/N$.

In the nutshell stochastic stability tell us that the Hamiltonian $H$ does not has any special 
features and that it properties are quite similar to those of similar random systems.

Although it seems quite natural, stochastic stability has quite deep consequences.  For example we 
could consider a system in which there are many equilibrium states, labeled by $\al$, and the 
overlaps among two generic configurations of the same state and of two different states are 
respectively $q_{M}$ and $q_{m}$, the free energies of the different states are uncorrelated\ldots 
The situation would be quite similar to the one described by one step replica symmetry breaking.  
However we may not specify the form of the probability distribution of the free energies which is 
characterized by a function ${\cal P}(f)$ which a priori may have an arbitrary shape.

It is a simple computation to verify that stochastic stability implies that the probability 
distribution of the free energies (${\cal P}(f)$) must have the form given in eq.  (\ref{REM}) with 
an appropriate choice of $m$.  The most dramatic effect of stochastic stability is to link the 
behaviour of the function ${\cal P}(f)$ in the region of large $f$ (where a large number of states 
do contribute) to the low $f$ behaviour, which controls the distribution of the states which are 
dominant in the partition function.

We have seen that stochastic stability strongly constraints the properties of the systems and many 
of the qualitative results of the replica approach can be derived as mere consequences of stochastic 
stability.  Stochastic stability apparently does not imply ultrametricity, which seems to be an 
independent property \cite{SOL}.  This independence  problem is still open as far has the only 
explicitly constructed probabilities distribution of the free energies of the states are 
ultrametric.

\subsection{Numerical  simulations}Let us consider the case of spin glasses at zero magnetic field (in this case the replica symmetry 
is fully broken \cite{BOOK}).  The natural variable to consider is a single spin ($A=\si_{i})$).  
In this case the correlation $C(t,t_{w})$ is equal to the overlap among two configurations at time 
$t$ and $t_{w}$:
\be
C(t,t_{w})={\sum_{i=N}\si_{i}(t)\si_{i}(t_{w})\over N}.
\ee
The response function is just the magnetization in presence of an infinitesimal magnetic field.
In this case the situation is quite good because there are reliable simulations for the system at 
equilibrium \cite{BOOK}.

In fig.  (1) (taken from \cite {MPRR}) we plot the prediction for the function $R$ versus $C$, 
obtained at equilibrium (i.e.  using the equilibrium probability distribution of the overlaps, 
$P(q)$) by means of a simulation of a $16^3$ lattice using parallel 
tempering~\cite{HUKUNEMOTO,BOOK}.  The simulation involves the study of $900$ samples of a $ L=16$ 
lattice.

During the off-equilibrium simulations \cite {MPRR} in a first run without magnetic field the 
autocorrelation function has been computed.  In a second second run from $t=0$ until $t=t_w$ the 
magnetic field is zero and then (for $t \ge t_w$) there is an uniform magnetic field of small 
strength $h_0$.  The starting configurations were always chosen at random (i.e.  the system is 
suddenly quenched from $T=\infty$ to the simulation temperature $T$).

In fig.  (1) there are the results of the off-equilibrium simulations \cite {MPRR} where $t_w=10^5$ 
and $t_w=10^4$, with a maximum time of $5\cdot 10^6$ Monte Carlo sweeps.  The lattice size in was 
$64$, and $T=0.7$ (well inside the spin glass phase, the critical temperature is close to 1.0).  We 
plot the response function $ R$ times $T$ (in this case $R$ is equal to $m/h_0$) against $C(t,t_w)$.  
We have plotted also a straight line with slope $-1$ in order to control where the FDT is satisfied.  
Finally we have plotted two points, in the left of the figure, that are obtained with the infinite 
time extrapolation of the magnetization.

The agreement among the absolute theoretical predictions (no free parameters) coming from the 
statics and the dynamical numerical data is quite remarkable.  These data show the correctness of the 
identification of the functions $x$ of the statics and $X$ of the dynamics.

Let us now go to the case of glass forming materials.  I will present the data for  binary 
mixture of soft spheres \cite{HANSEN}.  Theoretically there have been many speculations the glass 
transition is described by one step replica symmetry breaking \cite{KWT,PARI,MP,FRAPA}.  Here the 
equilibrium properties are no so well known as in spin glasses, although there are some evidence 
that the homologous of the function $P(q)$ is non trivial \cite{BAPA}.  On the other side, as we 
shall see, off-equilibrium simulations \cite{PAGE} show that the function $X(C)$ seems to be given 
by the one step formula (\ref{ONESTEP}) with an approximate linear dependence of $m$ on the 
temperature.

\begin{figure}
\epsfxsize=250pt\epsffile{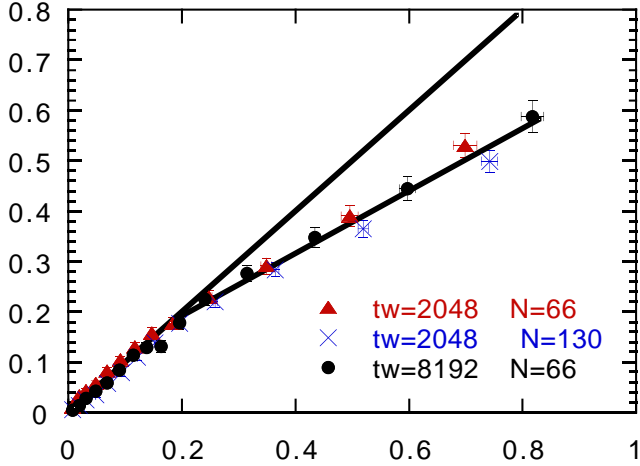}
\caption{ $R$ versus $\beta\Delta$ at $\Gamma=1.6$ for 
$t_{w}=8192$ and $t_{w}=2048$ at $N=66$ and for $t_{w}=2048$ at $N=130$.  The two straight lines 
have slope 1 and .62 respectively.}\label{UNO}
\end{figure}

We consider  a mixture of soft particles of different sizes.  Half of the particles 
are of type $A$, half of type $B$ and the interaction among the particles is given by the 
Hamiltonian:
\begin{equation}
H=\sum_{{i<k}} \left(\frac{(\si(i)+\si(k)}{|{\bf x}_{i}-{\bf x}_{k}|}\right)^{12},\label{HAMI}
\label{HAMILTONIAN}
\end{equation}
where the radius ($\si$) depends on the type of particles.  This model has been carefully studied in 
the past \cite{HANSEN,PAGE}.  The choice $\si_{B}/\si_{A}=1.2$ strongly inhibits crystallisation and 
the system goes into a glassy phase when it is cooled.  Using the same conventions of the previous 
investigators we consider particles of average radius $1$ at unit density.
It is usual to introduce the quantity $\Gamma 
\equiv \beta^{4}$.  For quenching from $T=\infty$ the glass transition is known to happen around
$\Gamma_c=1.45$
\cite{HANSEN}.

The best quantity we can measure to evidenziate off-equilibrium effects is the diffusion of the 
particles:
\be
\Delta(t,t_{w})\equiv {\sum_{i=1,N}\lan|{\bf x}_{i}(t_{w})-{\bf x}_{i}(t_{w}+t)|^{2} \ran \over N}.
\ee
The usual diffusion constant is given by $D=\lim_{t\to\infty}\Delta(t,t_{w})/t$.

The other quantity we measure is the response to a force.  At time $t_{w}$ we add to the Hamiltonian 
the term $\eps\ {\bf f} \cdot {\bf x}_{k}$, where $f$ is vector of squared length equal to $d=3$ and 
we measure the response
\be
R(t_{w},t)= {\partial \lan {\bf f} \cdot {\bf x}_{k}(t_{w}+t) \ran_{\eps}\over \partial \eps} {\Biggr 
|}_{\eps=0}
\approx { \lan {\bf f} \cdot {\bf x}_{k}(t_{w}+t)\ran_{\eps} \over \eps}
\ee
for sufficiently small $\eps$.
The usual fluctuation theorem tells that at equilibrium $\beta \Delta^{eq}(t)=R^{eq}(t)$. 

In the following we will look for the validity in the low temperature region of the generalized 
relation $\beta X(\Delta) =\partial R / \partial\Delta$.  This relation (with $X\ne 1$) can be valid 
only in the region where the diffusion constant $D$ is equal to zero.  Strictly speaking also in the 
glassy region $D\ne 0$, because diffusion may always happens by interchanging two nearby particles 
($D$ is different from zero also in a crystal); however if the times are not too large the value of 
$D$ is so small in the glassy phase that this process may be neglected in a first approximation.

The simulations we present are done using a Monte Carlo algorithm, which is a discretized form of a 
Langevin dynamics.  In fig.  \ref{UNO} we show $R$ versus $\beta\Delta$ at $t_{w}=2048$ and 
$t_{w}=8192$ for $\Gamma=1.6$ and $t\le 4t_{w}$ at $N=66$.  We also show the data for $t_{w}=2048$ 
at $N=130$.  We do not observe any significant systematic shift in this plot among three data sets.  
We distinguish two linear regions with different slope as expected from one step replica symmetry 
breaking.  The slope in the first region is compatible with 1, as expected from the FDT theorem, 
while the slope in the second region is near 0.62.  Also the data at different temperatures for all 
values of $\Gamma\ge 1.5$ show a similar behaviour.  The value of $R$, in the region where the FDT 
relation does not hold, can be very well fitted by a linear function of $\Delta$ as can be seen in 
fig.  \ref{UNO}.  The region where a linear fit (with $m<1$) is quite good corresponds to 
$t/t_w>0.2$.  

The fitted value of $m\equiv
\partial R/\partial (\beta
\Delta)$ is displayed in fig.  (3).  When $m$ becomes equal to 
1, the
fluctuation-dissipation theorem holds in the whole region and this is what happens at higher 
temperatures.  The straight line is the prediction of the approximation $m(T)=T/T_{c}$, using 
$\Gamma_{c}=1.45$.

All the results are in very good agreement with the theoretical expectations based on our knowledge 
extracted from the mean field theory for generalized spin glass models.  The approximation 
$m(T)=T/T_{c}$ seems to work with an embarrassing precision.  We can conclude that the ideas 
developed for generalized spin glasses have a much wider range of application than the models from 
which they have been extracted.  It likely that they reflect quite general properties of the phase 
space and therefore they can be applied in cases which are very different from the original ones.

\begin{figure}
\epsfxsize=250pt\epsffile{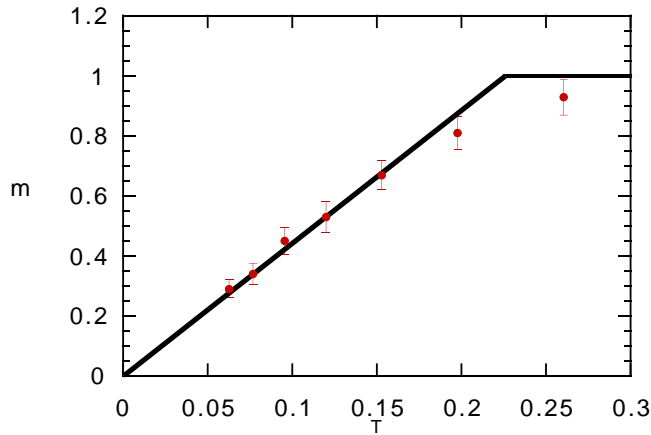}
\caption{ The quantity $m\equiv	{\partial R\over \partial \beta	\Delta}$ as $t_{w}=2048$ as
function of the temperature.  The straight line is the prediction of the approximation 
$m(T)=T/T_{c}$.}
\end{figure}\label{DUE}

\subsection{Some suggestions for planning an experiment}
The most interesting development would be to measure experimentally the function $X$ both in spin 
glasses and in structural glasses.  Clearly the most difficult task is the measurement of the 
fluctuations.  In spin glasses it is clear how it should be done: the measurement of the thermal 
fluctuations of the magnetization is a delicate, but feasible experiment.  In the case of structural 
glasses some ingenuity is needed in planning the experiments.  (A open interesting possibility would 
to do the measurements in the case of rubber, where a transition with similar characteristics should 
take place.)

\end{document}